# COATING OF ALUMINUM ALLOYS BY MICRO ARC OXIDATION IN NITRATE SALT

**Alexander Sobolev, Alexey Kossenko, Michael Zinigrad, Konstantin Borodianskiy**

*Zimin Advanced Materials Laboratory, Department of Chemical Engineering, Biotechnology and Materials, Ariel University, Ariel 40700, Israel*
*konstantinb@ariel.ac.il*

Plasma electrolytic oxidation (PEO) is a process for obtaining oxide coatings on valve metals. Mostly PEO is done in an aqueous solution electrolyte which limits the size of treated parts due to the system heating up. In presented work an alternative method of PEO processing applied in aluminum 1050 alloy in nitrate molten salt was investigated. The morphology, phase and chemical compositions, micro-hardness, and corrosion resistance were examined using. The obtained results showed that formed coating contains from two sub-layers, outer soft layer with the thickness of 4 µm and inner, denser layer with the thickness of 5µm. The formed coating consists of corundum, $\gamma$-$Al_2O_3$, $\theta$-$Al_2O_3$ and is free of any contaminants originated from the electrolyte.

## Introduction

Aluminum alloys show a variety of advanced physical and mechanical properties such as low density and high thermal and electrical conductivity [1]. Al alloys are widely applied in various industries such as electronics, marine, aerospace and machining [2]. However, its relatively low chemical resistance and negative potential limit its use, especially when good corrosion and wear resistance properties are required [3]. This issue could be solved by the formation of the protective coating formed by plasma electrolytic oxidation (PEO) process [4-6].

PEO process also named as micro arc oxidation (MAO) or anode spark deposition (ASD) is a type of high voltage anodizing process. The process causes to the formation of oxide crystal based coatings on valve metals and alloys (Al, Ti, Mg, Zr) [7-8].This process involves the reactions of anodizing and the formation of dielectric breakdowns (micro-arc discharges). While the voltage furthers increases, the micro-arc discharges stabilize and a crystalline high-temperature coating is forms. Usually the process conducted in aqueous electrolyte, but our previous works showed that molten salt electrolyte is preferable in applications for large surfaces [9-10] and in work of composition prediction [11].

In the presented work the process of coating formation on Al 1050 alloy by PEO process in nitrate salt of the $KNO_3 - NaNO_3$ composition was investigated. Electron microscopy and X-ray diffraction were applied to evaluate morphology, chemical and phase composition changes during the process. Microhardness and corrosion tests were performed to evaluate the technological properties of the formed coating.

### Experimental

Aluminum wrought alloy 1050 specimens (chemical composition shown in Table 1) with a surface area of 0.16-0.17 $dm^2$ were grounded using by abrasive papers grits #600-1000, and then subjected to ultrasonically cleaning in acetone.

Table 1

Chemical composition of the Al alloy 1050

| Chemical Element [%wt.] | | | | | | | |
|------|------|------|------|------|------|------|------|
| Si | Mg | Fe | Cu | Mn | Zn | Ti | Al |
| 0.25 | 0.05 | 0.44 | 0.05 | 0.05 | 0.07 | 0.05 | Base |

PEO treatment conducted at 280˚C in the electrolyte with the eutectic composition of $NaNO_3 - KNO_3$ with the mass % of 45.7:54.3 respectively. The electrolyte was held in a nickel crucible (99.95% Ni) which served as a counter electrode as well. The surface ratio of anode:cathode was 1:30, the anodic current density was 70 $mA/cm^2$, voltage was limited by the galvanostatic mode. The applied power supply has the following parameters: $I_{max} = 5A$, $U_{max} = 900V$, current and voltage were pulsed with a square-wave sweep at a frequency of 50Hz (ta = tk = 0.01 s). Duration of PEO treatment was 10min with the coating rate of 1μm/min.

The current and the voltage wave profiles and trend plots, as well as the power consumed during the process, were measured using Fluke Scope Meter 199C, which was located into the electrical circuit between the power supply and the working container.

A standard method for metallographic preparation was used for cross-sections preparation. Surface and cross-section morphologies of the obtained PEO coatings were examined by TESCAN MAIA3 scanning electron microscopy (SEM) equipped with an energy dispersive X-ray spectroscopy (EDS) system. The phase composition of the coatings were determined by a PANalytical Empyrean diffractometer (in Cu Kα radiation) in grazing incidence mode using a scan with a grazing angle of 3°, a step size of 0.03°, and at a 2θ range from 10° to 90°.

A Buehler Micromet 2100 microhardness tester was used to evaluate microhardness of the obtained oxide layer on a cross section area. The microhardness was determined according to ASTM E384, C1327, and B578 as the mean of eight measurements for each sublayer under a load equal to 10g.

The corrosion behavior of samples was examined by potentiodynamic polarization tests in a 3.5 wt.% NaCl solution by Autolab PGSTAT12 potentiostat/galvanostat with the General Purpose Electrochemical System (GPES), version 4.9 software. A three-electrode cell with a stainless steel counter electrode and a saturated Ag/AgCl reference electrode was used. The polarization resistance of a sample was determined at the range of ± 250mV with respect to the recorded corrosion potential at a scan rate of 0.1 mV/s (OCP). Prior to the potentiodynamic polarization test, the samples were kept in the 3.5 wt.% NaCl solution for 60 min in order to reach the steady state of a working electrode (OCP).

**Results and Discussion**

Fig. 1 presents the surface morphology of the treated by PEO alloy and the chemical composition obtained by EDS.

The obtained surface has a typical structure for the surfaces obtained by PEO approach. This coating has a pan-like structure with a nonuniformly distributed porous structure. The pores are of the sphere-like geometry with the average size of 0.2-2.5 µm. EDS results show that the results fit the composition of the formed aluminum oxide with no any additional impurities originated from the electrolyte composition. That is totally different from the results usually obtained in the aqueous electrolyte where the formed coating usually contains additional elements originated from the electrolyte composition.

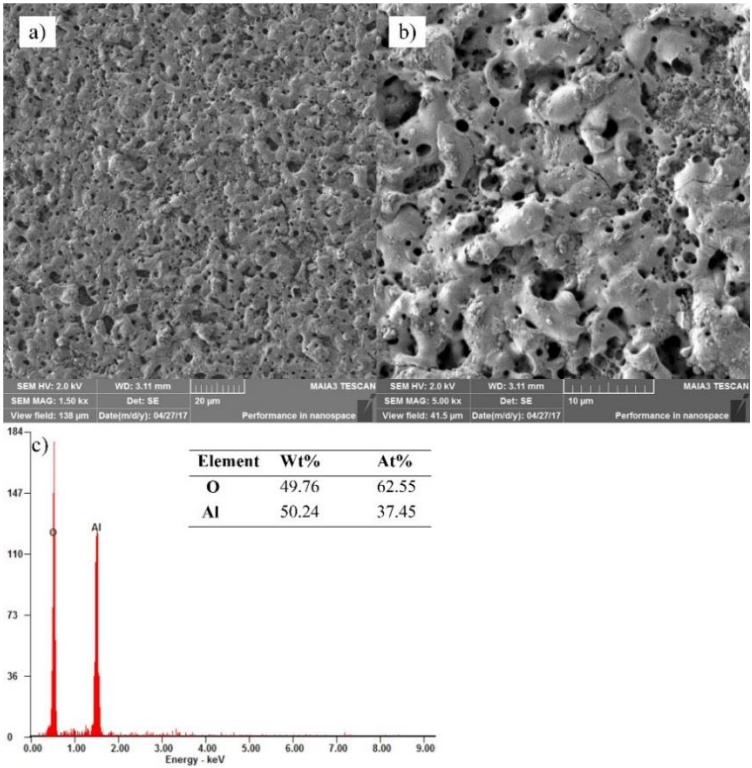

| Element | Wt% | At% |
|---------|-------|-------|
| O | 49.76 | 62.55 |
| Al | 50.24 | 37.45 |

**Fig. 1.** The surface morphology of Al 1050 alloy with oxide coating obtained by PEO approachshown in (a) ×1500, (b) ×5000, and (c) the chemical composition of the Al 1050 alloy after PEO treatment by EDS.

The line scan of the obtained coating is shown on Fig. 2.

The line scan of the obtained coating evaluated the presence of two layers: the outer and the inner. Based on the elemental distribution across the coating thickness it is seen that the inner layer is more homogeneous and has no pores while the outer has softer structure and contains pores. The thickness of the outer and the inner layers are about 4 and 5 µm, respectively.

The surface phase composition evaluation has been done by XRD investigation which patterns shown in Fig. 3.

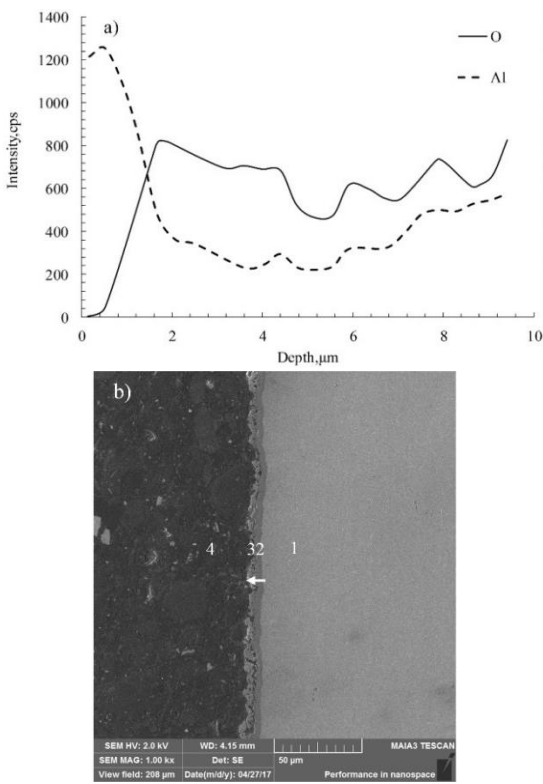

**Fig. 2.** Cross section morphology of Al 1050 alloy with oxide coating obtained by PEO approachand its EDS line scan: 1 – aluminum base (Al alloy 1050); 2 – inner layer; 3 –outer layer; 4 – resin.

XRD patterns detected the presence of three aluminum oxide phases namely, corundum, $\gamma$-$Al_2O_3$, $\theta$-$Al_2O_3$. Detected phases are phases of aluminum oxide which were formed as a result of high temperature dielectric plasma breakdowns with further melting and crystallization of newly formed coating.

Micro-hardness measurements were performed on the obtained oxide coating and compared with the base Al alloy. Those results are presented in Table 2. Based on the results, we assume that the outer layer mostly contains corundum phase while inner layer contains softer compounds as $\gamma$-$Al_2O_3$ and $\theta$-$Al_2O_3$.

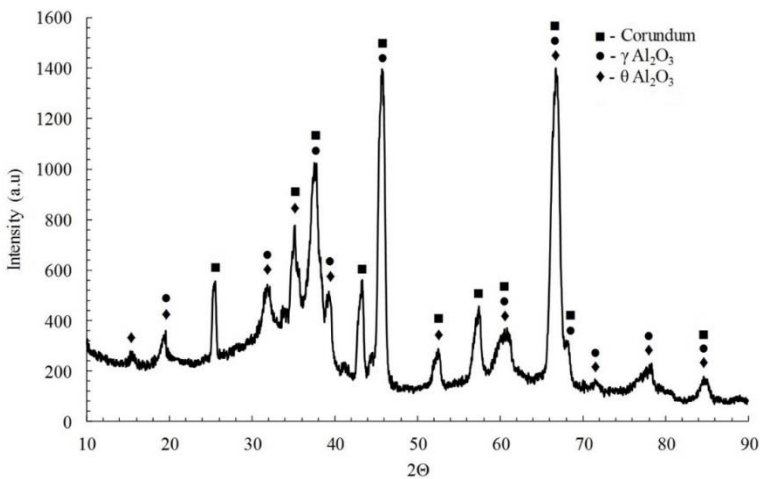

**Fig. 3.** X-ray surface diffraction pattern of Al 1050 alloy with oxide coating obtained by PEO approach.

Table 2

Micro-hardness measurementsthe alloy with oxide coating
obtained by PEO approach

| Base Al alloy [HV] | Inner layer[HV] | Outer layer[HV] |
|---|---|---|
| 44 | 748 | 1112 |

The corrosion resistance of the coated alloy was determined by the potentiodynamic polarization method. The obtained curve compared to that of unmodified ally and both are presented in Fig. 4.

Before the potenetiodynamic measurements, samples were immersed into 3.5% NaCl for reaching OCP. During the measurements the corrosion potential ($E_{corr}$) and the current ($i_{corr}$) were recorded. The coefficients $\beta_a$ and $\beta_c$ were calculated from the slops on the Figure 4 and the polarization resistance ($R_p$) was calculated according to the following equation:

$$R_p = \frac{\beta_a \times \beta_c}{2.3 \times i_{corr}(\beta_a + \beta_c)} \qquad (1)$$

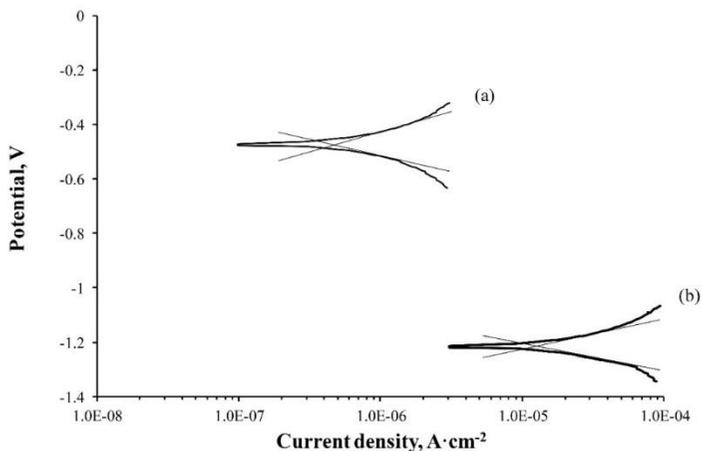

**Fig. 4**. Potentiodynamic polarization curves of Al 1050 alloy with oxide coating obtained by PEO approach (a) and base Al 1050 alloy (b), both in 3.5% NaCl solution.

Table 3 summarizes the obtained results on the corrosion parameters.

Table 3

Corrosion test characteristics of based Al 1050 alloy compared to the alloy with oxide coating obtained by PEO approach.

| Samples | Base Al 1050 | Coated Al 1050 |
|---|---|---|
| $E_{corr}$ [mV] | -980 | -480 |
| $i_{corr} \times 10^{-6}$ [A] | 13 | 0.5 |
| Anodic Tafel slope ($\beta_a$) [mV/decade] | 185 | 280 |
| CathodicTafel slope ($\beta_c$) [mV/decade] | 208 | 220 |
| $R_p, \times 10^5$ [$\Omega/cm^2$] | 0.032 | 1.07 |

The results in Table 3 show that PEO process is a promising method for improvement of metal corrosion resistance [12-13].The calculated polarization resistance of the coated alloy is $1.07 \cdot 10^5$ $\Omega/cm^2$, while uncoated is $0.032 \cdot 10^5$ $\Omega/cm^2$.Thepolarization resistance in coated alloy is one and a half orders of magnitude higher than that of the uncoated alloy.

**Conclusions**

The presented work showed the process and characterization of oxide coating formation on Al 1050 alloy in the eutectic composition nitrate salts. X-ray diffraction investigation results showed that the coating consists of the following phases: corundum, $\gamma$-$Al_2O_3$, $\theta$-$Al_2O_3$. The SEM images showed that the formed oxide coating contains two sublayers, outer soft layer with the thickness of 4µm and inner, denser layer with the thickness of 5µm. Those results were confirmed by micro-hardness measurements which detected the differences of the hardness of both oxide layers. Moreover, the corrosion resistance test showed that the corrosion rate of the coated alloy is in one order of magnitude lower compared to the uncoated alloy.

**Acknowledgements**

This work was carried out with the support of the Ministry of Aliyah and Integration, the State of Israel. Additionally, the authors would like to thank Ms. Natalia Litvak from the Engineering and Technology Unit at the Ariel University for her help with electron microscopy investigations.